\documentstyle[12pt]{article}
\newcommand{\A} {{\cal A}}

\title{OBSERVATION OF FSS FOR A  FIRST ORDER
PHASE TRANSITION}
\author{A. Billoire \\
Service de Physique Th\'eorique de Saclay\thanks{
Laboratoire de la Direction des Sciences de la Mati\`ere du CEA.} \\
91191 Gif-sur-Yvette Cedex, France      \\
T. Neuhaus\\
Universit\"at Bielefeld \\
Fakult\"at f\"ur Physik, Universit\"at Bielefeld,\\
D-W 4800 Bielefeld, FRG\\
and B. Berg \\
Wissenschaftskolleg zu Berlin\thanks{On sabbatical from the Florida
State University.}
\\ Wallotstr. 19, D-1000 Berlin 33, FRG  }
\date{November 3 1992}
\begin{document}
\maketitle

We present the results of a multicanonical simulation of the q=20 2-d Potts
model in the transition region. This is a very strong first order
phase transition. We observe, for the first time, the asymptotic
finite size scaling behavior predicted by Borgs and Koteck\'y
close to a first order phase transition point.

\vskip 2.cm
\rightline {SPhT-92/120}
\vfill\eject

\section{INTRODUCTION}

This paper addresses the question of lattice sizes needed in practical
applications in order to fulfil the asymptotic scaling behaviour
predicted by Borgs and Koteck\'y  \cite{Borgs_deux,Borgs_Miracle},
close to a first order phase transition point.

$\bullet$ Very strong first order transitions are easy to detect.
The system behavior is very close to the infinite volume behavior.
Ergodicity is broken. Thermodynamical quantities are discontinuous
at the transition point, with metastable branches.
A starting configuration half ordered, half disordered will relax to
very different states on both sides of the transition.

$\bullet$ In less clear cases, one must simulate systems of increasing
volumes $L^d$ and try to convince oneself that the above described very
large volume behavior is
approached. Let us use  the language of energy driven transitions in
what follows, and introduce the energy probability distribution
$P_L(E)$. In the transition region, it has two
peaks of heights $P^o_L$ (ordered phase) and $P^d_L$ (disordered phase),
separated by a minimum of height $P^{min}_L$.
For a first order transition, at fixed $P^o_L/P^d_L$,
$  {P^{min}_L /  P^o_L}$ decreases as $L$ grows.
This has been proposed by Lee and Kosterlitz \cite{Kosterlitz_two} as
an indicator of first order phase
transitions (for a microcanical version see  \cite{Brown,Gerling}).
One can also plainly look at a plot of $P_L(E)$ and decide ``by eye''
whether it looks more and more like two delta functions as $L$ grows.
Another class of indicators are moments of the distribution $P_L(E)$, that
go to zero in the large volume limit, for all temperatures,
but at a first order transition point. An example is the energy fluctuation
$CV/L^d=\beta^2  ( <E^2> - <E>^2 )$, another is Binder's
cumulant  $BL={1\over 3}(1-<E^4> / <E^2>^2)$.

$\bullet$ The most sophisticated (and trustworthy) method
is finite size scaling. One insists in seeing the  finite size behavior
as predicted by the theory in the vicinity of a first order phase transition.
In that case only one can be pretty sure that the trend observed for lattices
of increasing size does  continue up to the thermodynamical limit.
One  insists on seeing
\begin{equation}
 {1\over L^d} \ln{P^{min}(L)\over P^o(L)} \sim A / L
\end{equation}
\begin{equation}
CV_{max}/L^d  \sim CV^{(1)} + CV^{(2)} / L^d
\label{CV}
\end{equation}
\begin{equation}
BL_{min}  \sim BL^{(1)} + BL^{(2)} / L^d
\label{BL}
\end{equation}
where $CV_{max}$ is the maximum of the specific heat,
and $BL_{min}$ is the minimum of $BL$.

The above described large volume behavior has never really been
observed in a numerical simulation. The results of all
high precision simulations show
strong curvatures in e.g the curve representing $BL_{min}$ as a function of
$1/L^d$. This is the case of the 3-d $Z(3)$ Potts model with
antiferromagnetic admixture on lattices up to $48^3$ \cite{Paper_Z3}. This is
also the case of the 5 states 2-d Potts model \cite{Ukawa} on
lattices up to $256^2$. In the later case, the exact asymptotic limit is
$BL_{min} \sim -.44 \ 10^{-3}$ whereas the data presented would
suggest a much lower value ($\sim - 1.2 \ 10^{-3}$). This
shows that the $BL_{min}$ indicator is not a
trustworthy indicator of first order phase transitions on lattices
up to $256^2$ when the correlation length is equal to a few
thousands lattice spacings.

Results of a simulation of the $q=10$ 2-d Potts model on lattices
as large as $50^2$ can be found in Ref.  \cite{Paper_q10}. This
is a model with a strong, obvious, first order transition, with
an exactly known \cite{BW} correlation
length\footnote{The correlation length
of the 2-d Potts model at the transition point can be computed exactly
using the 6-vertex representation. The result is interpreted as the
disordered phase correlation length in the limit $\beta \to \beta_t^-$.
The ordered phase correlation length is smaller. Duality does not
relate theses two correlation lengths.}
$\xi(\beta_t^-) =10.56$.
The  results for the extrema of $CV/L^d$,
$BL$ and $U_4$ \cite{Paper_q10}
\begin{equation}
U4 =  {<(E-<E>)^4> \over{ <(E-<E>)^2>^2}}
\label{U4}
\end{equation}
and for the value $CV(\beta_t)/L^d$
are compared with the large volume predictions, namely
the extrema behave  like
$X^{(1)}+X^{(2)}/L^d+{\cal O}(1/L^{2d})$, and
$CV(\beta_t)/L^d$ behaves like
$X^{(1)}+X^{(2)}/L^d + {\cal O}(e^{-b L})$ for some $b > 0$.
The four different constants $\{X^{(1)}\}$ are exactly known. One
single unknown  parameter, e.g. the ordered specific
heat at the transition point $C_o$, determines the four $X^{(2)}$'s.
The precision of the data is very good, and
in all cases strong
deviations from the $X^{(1)}+X^{(2)}/L^d$ limiting behavior are seen.
These corrections do not seem to behave simply as
function of $L$, and are definitely not under control.
Not surprisingly, the values for the four slopes $\{X^{(2)}\}$ one would
infer from the
data give inconsistent estimates\footnote{Due to a systematic typing error,
every
times a numerical value  for  $C_d$ is given in Ref. \cite{Paper_q10}
(e.g. in Table~1), the value
for the ordered phase $C_o$ is meant.
In particular, the $q=10$  curves are drawn using the estimate of
$C_o=12.7 \pm .3$, not $C_d=12.7 \pm .3$.} of $C_o$.
Furthermore, the data do not substantiate the
prediction  that $CV(\beta_t)$ is asymptotic earlier than $CV_{max}$.

In conclusion, the FSS behavior is not observed even for a transition as
strong as the 2-d $q=10$ Potts model transition, on lattices as
large as $50^2$.
Three possible explanations come to the mind:
i) Eqs.\ref{CV},\ref{BL} are only proven in the large q limit, they may
not hold down to  $q=10$.
ii) Much larger lattices may be needed in order to
extract the true asymptotic behavior, even though on the largest lattices
simulated $P_L(E)$ has a textbook first order shape.
iii)~A programming error may be possible. We are however in the position
to compare high precision results for the $q=10$ Potts model, which were
obtained in two completely independent
simulations \cite{Paper_q10,Berg_Neuhaus}, e.g. using
different random number sequences. Results for thermodynamic quantities
agree on same lattice sizes to high precision, making the existence
of a programming error unlikely.

\section{Exact Results}

Those have been obtained  \cite{Borgs_deux} for models that
can be represented by a
contour expansion with small activities, like \cite{Borgs_Miracle}
the q states Potts
model for large q. In such a case,  the partition function for a
$L^d$ lattice with periodic boundary conditions
can be written as
\begin{equation}
 Z(\beta,L)  =   e^{- L^d \beta f_d(\beta)} + q e^{- L^d \beta f_o(\beta)}
 +   {\cal O}(e^{-b L})  e^{- \beta f(\beta) L^d }
\hskip .2cm ;\  b > 0
\label{Z}
\end{equation}
where $f_o(\beta)$ and $f_d(\beta)$ are smooth $L$ independent
functions. The free energy is $f(\beta) =
\min \{ f_o(\beta) , f_d(\beta) \}$.
The phenomenological two gaussian peak model of the energy
probability distribution $P_L(E)$ introduced by  Binder and
 Landau  \cite{Bind_un,Bind2}
follows through inverse Laplace transform. The above exact result fixes
the relative weights of the two peaks: At the infinite volume limit
transition point, $\beta = \beta_t$, the
ordered and disordered peak weights are exactly in the ratio q to one.

To the order in $1/L^d$ which we consider, all quantities are expressed
in terms of
$\beta_t$ and of the energies and specific heats of the two coexisting
phases. The transition temperature, $E_o$, $E_d$ and
the difference $C_o-C_d$  are known
exactly for the 2-d Potts models  \cite{Baxter}. Values for the
$q=10$ and $q=20$ cases can be found in Tab.\ref{Data}.
It follows from Eq.\ref{Z} that the specific heat
\begin{equation}
CV=\beta^2 L^d ( <E^2> - <E>^2 )
\end{equation}
has a maximum at
\begin{equation}
\beta(CV_{max}) = \beta_t - {\ln q \over{E_d-E_o}}{1\over{L^d}}
+{\beta^{(2)}_{CV} \over L^{2 d}} + {\cal O} (1/L^{3 d}).
\end{equation}

The height of this maximum increases linearly with $L^d$
\begin{equation}
CV_{max} = L^d  {{\beta_t^2}\over 4} (E_o - E_d)^2
+ CV^{(2)} + {\cal O} (1/{L^{d}}).
\end{equation}
whereas for fixed $\beta \neq \beta_t$, $CV(\beta)$ goes to a
constant, as $L$ goes to infinity.
One finds that $BL$ reaches a minimum equal
to  \cite{Comment,Borgs_deux,Kosterlitz}
\begin{equation}
BL_{min} =  -  {(E_o^2 - E_d^2)^2 \over {12 (E_o E_d)^2}}
+ {BL^{(2)} \over L^{d}} + {\cal O} (1/L^{2d})
\end{equation}
at the point
\begin{equation}
\beta(BL_{min}) = \beta_t -{\ln \bigl(q(E_o/E_d)^2\bigr)
\over{E_d-E_o}}{1\over L^d}
+{\beta^{(2)}_{BL} \over L^{2 d}} + {\cal O} (1/L^{3 d}).
\end{equation}

Expressions of the coefficients $\beta^{(2)}_{BL}$,  $BL^{(2)}$,
$\beta^{(2)}_{CV}$ and $CV^{(2)}$ as functions
of the $E_i$'s and $C_i$'s can be found in  \cite{Kosterlitz}.
$U4$ reaches a minimum \cite{Paper_q10}
\begin{equation}
U4_{min} =    1 + { 8 (C_o + C_d) \over{ L^d  \beta_t^2  (E_o-E_d)^2}}
+ {\cal O} (1/L^{2 d}).
\end{equation}
at the point
\begin{equation}
\beta(U4_{min}) = \beta_t -{\ln q
\over{E_d-E_o}}{1\over L^d}
+ {(C_o-C_d) (\ln^2q-8)\over{ L^{2 d} 2 \beta_t^2 (E_o-E_d)^3}}
+ {\cal O} (1/L^{3 d}).
\end{equation}

The above formulae for the extrema $CV_{max},\ BL_{min},\
U4_{min}$ and the corresponding effective $\beta$'s
have higher power law corrections that may hide the asymptotic
behavior on lattices  that can be simulated. In contrast,
the expressions for bulk averages evaluated at the (infinite volume
limit) transition point
$\beta=\beta_t$ do not have power law corrections,
as a consequence of Eq.\ref{Z}, e.g.  the specific heat behaves like
\begin{equation}
CV(\beta_t) = { C_d + C_o q   \over { 1 + q }}
+ {L^d q \over (1+q)^2} (E_o-E_d)^2 \beta_t^2
+ {\cal O}( e^{-b L})
\end{equation}

\section{Simulation of the q=20 model using the
Multicanonical Algorithm}

We have decided to simulate the $q=20$ Potts model,
as an example of a model with  much stronger phase transition
than the $q=10$ model (see Tab.1).
The conventional Metropolis (and Swendsen-Wang  \cite{Birosa})
algorithm suffers from exponential slowing down close to a strong
first order point, namely
the probability to jump from one phase to another
goes like $P_{o,d} \sim e^{-2 \A_{o,d} L^{d-1}}$,
where  $\A^{od}$ is the order-disorder surface tension. This makes
simulation of a $\approx \ 50^2$ lattice with such algorithms
impossible even with vastly more powerful computers than available now.

\begin{table} [h]
\begin{small}
\begin{center}\begin{tabular}{|lrrrrc|}\hline
q   &$C_o$             & $E_d-E_o$   & $C_d-C_o$  & $\xi_{\beta_c^-}$ &
${(C_o/(E_o-E_d)^2)}^{1/d}$\\
[3pt]\hline
10  &$12.7 \pm .3$     & 0.69605 & .44763    &  10.56 & $5.12 \pm .06$ 	     \\
20  &$5.2  \pm .2$     & 1.19416 & .77139    &   2.70 & $1.91  \pm .04$	\\
[3pt]\hline
\end{tabular}\end{center}
\end{small}
\label{Data}
\caption{The $q=10$ and $q=20$ 2-d Potts model: Estimated values of $C_o$,
exact values for $E_d-E_o$, $C_d-C_o$, and correlation length
$\xi(\beta \to \beta_t^-$). A transition is strongly first order, on
$L^d$ lattice  when $L >> \xi$, $L^d >> C_o /(E_d-E_o)^2$}
\end{table}

Such a simulation became possible with the invention of the
``multicanonical ensemble'' \cite{Berg_Neuhaus}. It amounts to
perform the simulation with a weight factors designed in such a way
that $P_L(E)$ is smooth
between $E_o$ and $E_d$, and to ``reweight'' the events when computing
expectation values. The   multicanonical
algorithm has only polynomial slowing down.

\begin{table} [h]
\begin{small}
\begin{center}\begin{tabular}{|lrr|}\hline
$L$       &  M   sweeps      &        bins          \\
[3pt]\hline
16        &             3.7  &         8 \\
16        &            10.2  &	      20 \\
16        &            90.5  &	     180\\
[3pt]\hline
18	  &	        10.0 & 	      20\\
18	  &	        10.0 &	      20\\
18	  &	        10.0 &	      20\\
18	  &             30.0 &	      60\\
[3pt]\hline
20	  &             60.0 & 	     120\\
[3pt]\hline
24        &              7.7 &        15\\
24        &             10.2 &	      20\\
24        &             50.5 &	     100\\
24        &             20.2 &	      40\\
24        &             40.5 &	      80\\
[3pt]\hline
30        &             10.2 & 	      20\\
30	  &	        10.0 &	      20\\
30        &             50.5 &	     100\\
[3pt]\hline
32	  &             40.5 & 	      80\\
[3pt]\hline
34        &             7.2  &        14\\
34        &            29.7  &	      59\\
[3pt]\hline
38	  &             40.  &        80\\
38	  &	        10.  &	      20\\
38	  & 	        10.  &	      20\\
\hline
\end{tabular}\end{center}
\end{small}
\label{Runs}
\caption{ $q=20$ 2-d Potts model: Simulated lattices.
Lattice sizes,  number of mega sweeps
performed  and number of bins used to perform the analysis.
The first bin may contain a different number of sweeps that the others.
It has always been discarded  for thermalization.}
\end{table}

\begin{table} [h]
\begin{small}
\begin{center}\begin{tabular}{|l|rrrr|}\hline
L & $CV_{max}/L^2$ & $BL_{min}$ & $U4_{min}$ & $CV(\beta_t)/L^2$ \\
[3pt]\hline
16 & 1.03647 (15 ) & -0.59302 (15 ) & 1.07660 (6 ) & 0.2066 (12 ) \\
18 & 1.03598 (17 ) & -0.58489 (17 ) & 1.06186 (5 ) & 0.2010 (16 ) \\
20 & 1.03582 (19 ) & -0.57878 (18 ) & 1.05080 (5 ) & 0.1988 (22 ) \\
24 & 1.03482 (10 ) & -0.57013 (10 ) & 1.03602 (2 ) & 0.1924 (17 ) \\
30 & 1.03360 (13 ) & -0.56230 (14 ) & 1.02352 (2 ) & 0.1952 (34 ) \\
32 & 1.03317 (22 ) & -0.56058 (24 ) & 1.02080 (3 ) & 0.1923 (58 ) \\
34 & 1.03286 (18 ) & -0.55911 (17 ) & 1.01851 (4 ) & 0.1837 (55 ) \\
38 & 1.03218 (12 ) & -0.55653 (12 ) & 1.01492 (2 ) & 0.1941 (56 ) \\
\hline
\end{tabular}\end{center}
\end{small}
\label{Results}
\caption{  $q=20$ 2-d Potts model: Results for $CV_{max}/L^2$, $BL_{min}$,
$U4_{min}$ and $CV(\beta_t)/L^2$.}
\end{table}

The multicanonical algorithm is only partially vectorizable.
On a $38^2$ lattice, our program takes 7.7 micro-sec to update a spin
on a CRAY X-MP, 27.8 micro-sec on a RS6000-550 and 35.0 micro-sec
on a DEC 6000-510.
We have made runs on lattices indicated in Tab.\ref{Runs}.
Most runs were performed on workstations, using the
pseudo random number sequence $I_i=I_{i-147}\otimes I_{i-250}$, where $\otimes$
is the logical exclusive OR. Run $\#4$ at $\L=24$, and run $\#1$
at $L=38$ were performed on a CRAY X-MP, using the congruential
pseudo random number generator RANF().
Statistical analysis was performed using Jackknife technique, with
first order bias correction.
In many cases, we performed several simulations for a given
lattice size, using more or less optimized weights.
Each run gave us results for e.g. $CV_{max}$. Those were averaged
using weighted averages. In some  cases, we obtained
$\chi^2$ as bad as $\approx 2 \ p.d.f$. We believe this
would disappear with higher statistics.

Our results for $CV_{max}/L^d$, $BL_{min}$ and $U4_{min}$
can be found in Figs.\ref{Fig.CV_max}, \ref{Fig.BL_min} and \ref{Fig.U4_min}
together with the theoretical estimate
using the value $C_o = 5.2 \pm .2$. In all cases the predicted
$X^{(0)} + X^{(1)} / L^d$ behavior is approached on the
largest lattices, furthermore the
values of the three slopes $\{X^{(2)}\}$ one infer from our
data give consistent estimates of $C_o$.
This was not true in case of the $q=10$ simulation on similar size lattices.
Our results for $\beta(CV_{max})$ and $\beta(BL_{min})$
can be found in Fig.\ref{Fig.betas}. Perfect agreement is found
with the theoretical predictions. The effect of the $1/L^{2d}$ correction
is not visible, and the data for $\beta(U4_{min})$ (not plotted)
are not distinguishable from those for  $\beta(CV_{max})$.
The behavior of $CV(\beta_t)$
is shown in Fig.\ref{Fig.CV_beta}. No deviation is seen
from the predicted $X^{(0)} + X^{(1)} / L^d$ behavior.
This agrees with the prediction that corrections
are ${\cal O}(e^{-bL})$.

In conclusion our data for the $q=20$ Potts model are in excellent
agreement with Finite Size Scaling predictions. In particular the
data for  $CV(\beta_t)$ agree with FSS for all lattices considered.

This indicates  that  the disagreement between FSS
and data for the 2-d $q=10$ Potts model were small system effects.
It means however than
the asymptotic behavior predicted by Binder, and
later proven by  Borgs and  Koteck\'y, only sets in for very large
lattices.
The lattice size must fulfil the conditions $L >> \xi$, $L^{d-1} >>
1/\A^{od}$
(If Widom's relation  \cite{Widom} holds this condition is equivalent
to the first one),
and $L^d >> C_o / (E_o-E_d)^2$, $L^d >> C_d / (E_o-E_d)^2$,
where $>>$ means five to ten
times larger. It is unfortunate that for such large systems, the
transition is already blatantly first order.

\begin{figure} [tbh]
\vglue  10.5cm
\includegraphics{Fig1.ps}
\caption{ $CV_{max}/L^2$ as a function of $1/L^2$ for the 2-d q=20
Potts model. The three curves are drawn using the central value and
the one standard estimates for $C_o$.}
\label{Fig.CV_max}
\end{figure}

\begin{figure} [tbh]
\vfill\penalty -5000\vglue 8.5cm
\includegraphics{Fig3.ps}
\caption{$BL_{min}$ as a function of $1/L^2$ for the 2-d q=20
Potts model. The three curves are drawn using the central value and
the one standard estimates for $C_o$.}
\label{Fig.BL_min}
\end{figure}

\begin{figure} [tbh]
\vfill\penalty -5000\vglue 8.5cm
\includegraphics{Fig5.ps}
\caption{$U4_{min}$ as a function of $1/L^2$ for the 2-d q=20
Potts model. The three curves are drawn using the central value and
the one standard estimates for $C_o$.}
\label{Fig.U4_min}
\end{figure}

\begin{figure} [tbh]
\vfill\penalty -5000\vglue 8.5cm
\includegraphics{Fig6.ps}
\caption{$\beta(CV_{max})$ and $\beta(BL_{min})$  as a
function of $1/L^2$ for the 2-d q=20
Potts model. The  curves are drawn using the central value
for $C_o$. The effect of the uncertainty on $C_o$ would not be visible.}
\label{Fig.betas}
\end{figure}

\begin{figure} [tbh]
\vfill\penalty -5000\vglue 8.5cm
\includegraphics{Fig10.ps}
\caption{$CV(\beta_t)/L^2$ as a function of $1/L^2$ for the 2-d q=20
Potts model. The three curves are drawn using the central value and
the one standard estimates for $C_o$.}
\label{Fig.CV_beta}
\end{figure}

\section*{Acknowledgements}

This work was in part supported by the
U.S. Department of Energy under contracts DE-FG05-87ER40319 and
DE-FC05-85ER2500.

\end{document}